\documentclass[12pt,final]{elsart}

  \usepackage{latexsym,bm,amsmath,amssymb,amsfonts}
  \usepackage{epsfig,graphics,graphicx}
  \usepackage{slashed}
\usepackage{latexsym,bm,amsmath,amssymb,amsfonts}

\newcommand{\beq}{\begin{eqnarray}}
\newcommand{\eeq}{\end{eqnarray}}

\begin{document}

\begin{frontmatter}

\parbox[]{16.0cm}{ \begin{center}
\title{Notes on the orbital angular momentum of quarks \\ in the nucleon}

\author{Yoshitaka Hatta  }

\address{  Faculty of Pure and Applied Sciences, University
of Tsukuba, \\Tsukuba, Ibaraki 305-8571, Japan
}

\end{center}


\begin{abstract}
We discuss the  orbital angular momentum of partons inside a longitudinally polarized proton  in the recently proposed framework of  spin decomposition. The quark orbital angular momentum defined by Ji can be decomposed into the `canonical' and the `potential' angular momentum parts, both of which are represented as the matrix element of a manifestly gauge invariant operator.
\end{abstract}
}

\end{frontmatter}

\vspace{10mm}

\section{Introduction}

Recent polarized beam experiments and global QCD analyses suggest that the contribution of the gluon helicity $\Delta G$ to the
spin of the proton is rather small \cite{Boer:2011fh}. This observation, together with the inexorable fact that the quark helicity contribution $\Delta \Sigma$ is also small (less than $30\%$), lead one to suspect that the key to understand the proton spin puzzle is the orbital angular momentum (OAM) of quarks and gluons. However, progress in this direction has been hindered by a number of difficulties in measuring, and even defining the OAM.
  So far, the only well--recognized, gauge invariant definition of the quark OAM  is the one by Ji \cite{Ji:1996ek}
   which can be measured, indirectly, as the difference between a certain moment of the generalized parton distribution and  $\Delta \Sigma$. Although generally accepted, this approach may be criticized on the basis that the corresponding operator is not the `canonical' one that satisfies the fundamental commutation relation of the angular momentum operator in quantum mechanics. Efforts to improve upon this point have led Chen {\it et al.} to propose a completely new decomposition scheme of the QCD angular momentum tensor \cite{Chen:2008ag,Chen:2009mr} which has triggered a  flurry of activity lately
   \cite{Wakamatsu:2010qj,Wakamatsu:2010cb,Cho:2010cw,Chen:2011gn,Leader:2011za,Hatta:2011zs,Wakamatsu:2011mb,Chen:2011zzh,Zhang:2011rn,Lin:2011us}.
    However, the issue still remains very controversial, and the overarching impact of this new formalism as well as its practical usefulness in phenomenology are yet to be clarified.

In this work, we investigate the quark OAM along the line of our previous work \cite{Hatta:2011zs} which we view as the proper rendition of the formalism \cite{Chen:2008ag,Chen:2009mr} in the context of high energy QCD. We shall show that one can represent the canonical OAM as the matrix element of a manifestly gauge invariant operator which turns out to be equivalent to that obtained in the Wigner distribution approach  \cite{Lorce:2011kd}.  This paves the way to measure the canonical OAM experimentally or numerically on a lattice, and thus helps to mitigate the criticism that the matrix elements defined in \cite{Chen:2008ag} do not have known physical measurements \cite{Ji:2010zza}.  Actually, in the gluon helicity sector Ref.~\cite{Hatta:2011zs} has already shown how one can reconcile the gluon helicity defined in  \cite{Chen:2008ag} with $\Delta G$ which is measurable.
We now extend this finding to the OAM sector.

\section{Decomposition of the QCD angular momentum operator }

The main idea of \cite{Chen:2008ag,Chen:2009mr} is that one can achieve a complete, gauge invariant decomposition of the QCD angular momentum operator by identifying the `physical' and `pure gauge' components of the gauge field
\beq
&& A^\mu = A^\mu_{\scriptsize\mbox{phys}} + A^\mu_{\scriptsize\mbox{pure}}\,, \\
&& F_{\scriptsize\mbox{pure}}^{\mu\nu}
= \partial^\mu A^\nu_{\scriptsize\mbox{pure}} -\partial^\nu A^\mu_{\scriptsize\mbox{pure}} +ig [A^\mu_{\scriptsize\mbox{pure}},A^\nu_{\scriptsize\mbox{pure}}] = 0\,,  \label{pure}
\eeq
 which transform differently under gauge transformations
 \beq
 && A^\mu_{\scriptsize\mbox{phys}} \to U^\dagger A^\mu_{\scriptsize\mbox{phys}}U\,, \nonumber \\
 && A^\mu_{\scriptsize\mbox{pure}} \to U^\dagger A^\mu_{\scriptsize\mbox{pure}} U -\frac{i}{g} U^\dagger \partial^\mu U\,. \label{tran}
 \eeq
  The QCD angular momentum tensor $M^{\mu\nu\lambda}$ can then be written as a sum of the helicity and the  orbital angular momentum of quarks and gluons. In the `covariant' form  \cite{Wakamatsu:2010qj,Wakamatsu:2010cb}
  useful for high energy experiments, the original proposal by Chen {\it et al}.  \cite{Chen:2008ag,Chen:2009mr}  reads
 \beq
 M_{\scriptsize\mbox{quark-spin}}^{\mu\nu\lambda} &=& -\frac{1}{2}\epsilon^{\mu\nu\lambda\sigma}\bar{\psi}
\gamma_5 \gamma_\sigma  \psi\,, \\
 M_{\scriptsize\mbox{quark-orbit}}^{\mu\nu\lambda}&=&\bar{\psi}\gamma^\mu (x^\nu iD_{\scriptsize\mbox{pure}}^\lambda
-x^\lambda iD_{\scriptsize\mbox{pure}}^\nu )\psi\,,  \label{24} \\
 M_{\scriptsize\mbox{gluon-spin}}^{\mu\nu\lambda}&=&  F_a^{\mu\lambda}A_{\scriptsize\mbox{phys}}^{\nu a} -
F_a^{\mu\nu}A_{\scriptsize\mbox{phys}}^{\lambda a}  \,,  \label{glu}  \label{25} \\
 M_{\scriptsize\mbox{gluon-orbit}}^{\mu\nu\lambda}&=&  F_a^{\mu\alpha}\bigl(x^\nu (D_{\scriptsize\mbox{pure}}^\lambda A_\alpha^{\scriptsize\mbox{phys}})_a
-x^\lambda (D^\nu_{\scriptsize\mbox{pure}}A_\alpha^{\scriptsize\mbox{phys}})_a \bigr)\,. \label{26}
\eeq
 where $D^\nu_{\scriptsize\mbox{pure}} \equiv \partial^\nu +igA^\nu_{\scriptsize\mbox{pure}}$, and $a,b=1,2,\cdots,8$ are the color indices.\footnote{
Our convention is $
\epsilon^{0123}=+1\,, \gamma_5=-i\gamma^0\gamma^1\gamma^2\gamma^3$.
 We shall use the light--cone coordinates  $x^\pm = \frac{1}{\sqrt{2}}(x^0\pm x^3)$ and denote  the transverse coordinates with latin indices $x_T = \{x^i\}$, $(i,j,\cdots=1,2)$. The two--dimensional antisymmetric tensor is defined  as
$\epsilon^{ij}=-\epsilon^{+-ij}\,, \epsilon^{12}=\epsilon_{12}=-\epsilon^{21}=1$.
} Using the transformation rule (\ref{tran}), it is easy to check that each of the above components is gauge invariant. Such a complete decomposition goes beyond Ji's framework in which the gluonic part cannot be separated into the helicity and orbital parts. The price to pay, however, is that the decomposition is not local in the sense that   $A^\mu_{\scriptsize\mbox{phys}}$ is in general nonlocally related to the total $A^\mu$. Moreover, it is not entirely covariant, either, because $A^\mu_{\scriptsize\mbox{phys}}$ actually depends on the frame  as we shall soon see.
An alternative decomposition of the orbital part was suggested by Wakamatsu \cite{Wakamatsu:2010qj}
\beq
 M_{\scriptsize\mbox{quark-orbit}}^{'\mu\nu\lambda}&=&\bar{\psi}\gamma^\mu (x^\nu iD^\lambda
-x^\lambda iD^\nu )\psi\,,  \label{ji} \\
 M_{\scriptsize\mbox{gluon-orbit}}^{'\mu\nu\lambda}&=&  F_a^{\mu\alpha}\bigl(x^\nu (D_{\scriptsize\mbox{pure}}^\lambda A_\alpha^{\scriptsize\mbox{phys}})_a
-x^\lambda (D^\nu_{\scriptsize\mbox{pure}}A_\alpha^{\scriptsize\mbox{phys}})_a \bigr)  \nonumber \\ && \qquad \qquad \qquad \qquad
+ (D_\alpha F^{\alpha\mu})_a
(x^\nu A_{\scriptsize\mbox{phys}}^{\lambda a} -x^\lambda A_{\scriptsize\mbox{phys}}^{\nu a}) \,. \label{waka}
\eeq
The second term of (\ref{waka}) is gauge invariant on its own. Using the equation of motion $D_\alpha F^{\alpha \mu}_a = g\bar{\psi}\gamma^\mu t^a \psi$, one sees that it accounts for the difference between $D^\nu_{\scriptsize\mbox{pure}}$ in (\ref{24}) and  $D^\nu$ in (\ref{ji}).

There is no consensus as to which definition, (\ref{24}) or (\ref{ji}), is more appropriate for the quark orbital angular momentum. To some extent, it is a matter of choice. It is (\ref{24}), but not (\ref{ji}), that is compatible with the (equal--time) canonical commutation relation of the angular momentum operator  $\vec{L}\times \vec{L}=i\vec{L}$,
\beq
\vec{L}=\vec{x} \times i\vec{D}_{\scriptsize\mbox{pure}}\,, \qquad \vec{x}=(x^1,x^2,x^3)\,.
\eeq
The pure gauge condition (\ref{pure}) is crucial for this. (\ref{24}) may thus be called the canonical angular momentum.\footnote{Throughout this paper, we associate the term  `canonical' with the operator $iD^\mu_{\scriptsize\mbox{pure}}$ instead of the usual $i\partial^\mu$. The former is actually the gauge covariant generalization of the latter without affecting the commutation relation.}
On the other hand, (\ref{ji}) is the same as  Ji's definition \cite{Ji:1996ek} and is accessible from the analysis of the generalized parton distribution (GPD),  whereas it has not been known how to measure (\ref{24}).

 We will derive an explicit expression of the canonical angular momentum (\ref{24}) in terms of a  manifestly gauge invariant  operator whose matrix element is, in principle, related to experimental processes or observables in lattice QCD simulations. For this purpose, one must specify what $A^\mu_{\scriptsize\mbox{phys}}$ is. There are several proposals for $A^\mu_{\scriptsize\mbox{phys}}$ in the literature \cite{Chen:2009mr,Cho:2010cw,Hatta:2011zs,Zhang:2011rn}. These definitions are not equivalent as suggested by the work of Ref.~\cite{Chen:2011gn} which showed that  they give different  values of the gluon helicity (the proton matrix element of (\ref{25})). Here we employ the one proposed in \cite{Hatta:2011zs}
\beq
A^{\mu }_{\scriptsize\mbox{phys}}(x) = -\int dy^- {\mathcal K}(y-x) {\mathcal W}^{-}_{xy}
F^{+\mu}(y^-,\vec{x}){\mathcal W}^{-}_{yx}\,, \label{hatta}
\eeq
where we use the notation $\vec{x}=(x^+,x^i)$ from now on. ${\mathcal W}$ is the Wilson line operator
\beq
{\mathcal W}^{-}_{xy} \equiv {\mathcal P}
 \exp\left(-ig\int^{x^-}_{y^-} A^+(y'^-,\vec{x}) dy'^- \right)\,,
\eeq
  in the fundamental representation. The superscript `$-$' denotes that the path ordering is in the $x^-$ direction.
 ${\mathcal K}(y^-)$ is either $\frac{1}{2}\epsilon(y^-)$, $\theta(y^-)$ or $-\theta(-y^-)$, depending on  the boundary condition at $x^- = \pm \infty$ in the light--cone gauge $A^+=0$.\footnote{From the viewpoint of the $PT$ (parity and time--reversal) symmetry which will be crucially used below, it seems that the choice ${\mathcal K}(y^-)=\frac{1}{2}\epsilon(y^-)$ is the most natural and convenient one, although the difference does not matter in the end. }
The pure gauge part $A_{\scriptsize\mbox{pure}}$ is
\beq
A^\mu_{\scriptsize\mbox{pure}}(x)\equiv - \frac{i}{g}{\mathcal W}^{-}_{x,\pm\infty}{\mathcal W}_{\pm\infty} \partial^\mu ({\mathcal W}^{-}_{x,\pm\infty}{\mathcal W}_{\pm\infty})^\dagger\,, \label{pu}
\eeq
where ${\mathcal W}_{\pm\infty}={\mathcal P}
 \exp\left(-ig\int^{\vec{x}}_{\vec{\infty}} \vec{A}(\pm \infty,\vec{x}')\cdot d\vec{x}' \right)$ is the  Wilson line in the spatial direction at $x^- = \pm \infty$.
 It represents the residual gauge symmetry of the light--cone gauge $A^+=0$, and  is fixed by  specifying the boundary condition of the gauge field at $x^- \to \pm \infty$ mentioned above.
  It has been shown in \cite{Hatta:2011zs} that (\ref{hatta}) and (\ref{pu}) are a viable decomposition of the total gauge field $A^\mu$. We wish to stress that this particular choice  is singled out among others by the criterion of measurability: The corresponding gluon helicity  coincides with the usual gluon helicity $\Delta G$ that has been measured in experiments.

 Note that  the definition (\ref{hatta}) already selects a particular frame---the infinite momentum frame where the partonic interpretation of hadrons is clearest. As emphasized in \cite{Goldman:2011vs}, the decomposition of spin into the helicity and the orbital parts cannot be made entirely  covariant, but depends on the frame of reference.

\section{Potential  angular momentum}

We now focus on the orbital angular momentum of quarks inside a longitudinally polarized proton. It is given by the forward matrix element of the $\mu\nu\lambda=+ij$ component of  (\ref{24}) or (\ref{ji})
\beq
\epsilon^{ij}L_{\scriptsize\mbox{Chen}}&\equiv & \frac{1}{2P^+}\frac{ \langle PS| \int dx^-d^2x_T M_{\scriptsize\mbox{quark-orbit}}^{+ij}| PS\rangle }{(2\pi)^3\delta^{3}(0) }\nonumber \\
&=& \frac{1}{2P^+} \frac{\langle PS| \int dx^-d^2x_T \, \bar{\psi}\gamma^+ (x^i iD^j_{\scriptsize\mbox{pure}}- x^j iD_{\scriptsize\mbox{pure}}^i)\psi|PS\rangle }{ (2\pi)^3\delta^{3}(0)} \,, \label{jm}
\eeq
\beq
\epsilon^{ij}L_{\scriptsize\mbox{Ji}}  &\equiv & \frac{1}{2P^+}\frac{\langle PS|  \int dx^- d^2x_T M_{\scriptsize\mbox{quark-orbit}}^{'+ij}| PS\rangle}{(2\pi)^3\delta^{3}(0)} \nonumber \\
&=& \frac{1}{2P^+} \frac{\langle PS| \int dx^- d^2x_T \, \bar{\psi}\gamma^+ (x^i iD^j- x^j iD^i)\psi  |PS\rangle}{(2\pi)^3\delta^{3}(0)} \,,
\eeq
 where $P^2=-S^2=M^2$ (the proton mass squared) and $(2\pi)^3\delta^3(0)=\int dx^- d^2x_T$ is the momentum space delta function. The longitudinal polarization means $S^\mu =(S^+,S^-,S_T) \approx (S^+,0,0_T)$.
 We first observe that, since   $A^\mu_{\scriptsize\mbox{pure}}=0$ in the light--cone gauge $A^+=0$ \cite{Hatta:2011zs}, $L_{\scriptsize\mbox{Chen}}$ is actually identical to the Jaffe--Manohar (JM) definition \cite{Jaffe:1989jz} of the quark orbital angular momentum
 \beq
 L_{\scriptsize\mbox{JM}}=
\frac{1}{2P^+}\frac{\langle PS| \int dx^- d^2x_T\,  \bar{\psi}\gamma^+(x^1 i\partial^2-x^2 i\partial^1) \psi |PS\rangle_{\scriptsize\mbox{LC}}}{(2\pi)^3\delta^{3}(0)}\,.
\eeq
  The subscript ${\scriptsize\mbox{LC}}$ means that the matrix element is evaluated in the light--cone  gauge.
  In other words, $L_{\scriptsize\mbox{Chen}}$ is the generalization of $L_{\scriptsize\mbox{JM}}$ to arbitrary gauges  (see, also, \cite{Wakamatsu:2010qj}). We thus unify the notations  $L_{\scriptsize\mbox{Chen}} =L_{\scriptsize\mbox{JM}} \equiv L_{\scriptsize\mbox{can}}$ by introducing the canonical orbital angular momentum $L_{\scriptsize\mbox{can}}$, and write
 \beq
 L_{\scriptsize\mbox{Ji}} = L_{\scriptsize\mbox{can}} +L_{\scriptsize\mbox{pot}}\,,
 \eeq
  where the so--called potential angular momentum \cite{Burkardt:2008ua,Wakamatsu:2010qj} is, with our choice of
  $A^\mu_{\scriptsize\mbox{phys}}$,
\beq
\epsilon^{ij}L_{\scriptsize\mbox{pot}} &=&
\frac{1}{2P^+ (2\pi)^3\delta^{3}(0)}\langle PS| \int dx^- d^2x_T\, x^i\bar{\psi}(x) \gamma^+ (-g)(x^iA^j_{\scriptsize\mbox{phys}} -x^jA^i_{\scriptsize\mbox{phys}})\psi(x)|PS\rangle \nonumber \\
&=& \frac{1}{2P^+(2\pi)^3\delta^{3}(0) } \langle PS| \int dx^- d^2x_T\, \Bigl\{   x^i\bar{\psi}(x)
 \gamma^+ \int dy^- {\mathcal K}(y^- - x^-) {\mathcal W}^-_{xy}\,
gF^{+j}(y^-,\vec{x}){\mathcal W}^-_{yx} \psi(x)
  \nonumber \\
  && - x^j  \bar{\psi}(x) \gamma^+ \int dy^- {\mathcal K}(y^- - x^-) {\mathcal W}^-_{xy}\, g
F^{+i}(y^-,\vec{x}){\mathcal W}^-_{yx} \psi(x) \Bigr\} |PS\rangle  \,.
\label{phys}
\eeq
 Now consider the non-forward matrix element
\beq
 \frac{1}{2\bar{P}^+(2\pi)^3\delta^{3}(0)} \langle P'S'|\int dx^- d^2x_T\, x^i\bar{\psi}(x)
 \gamma^+
\int dy^- {\mathcal K}(y^- - x^-)\, {\mathcal W}^-_{xy} \, g F^{+j}(y^-,\vec{x}) {\mathcal W}^-_{yx} \psi(x)
 |PS\rangle\,, \nonumber
\eeq
 where $\bar{P}^\mu=(P^\mu+P'^\mu)/2$ and the momentum transfer will be denoted as  $\Delta^\mu \equiv P'^\mu-P^\mu$.
 The explicit factor $x^i$ can be traded for the
  derivative with respect to $\Delta^i$ such that
\beq
 &&  \epsilon^{ij}L_{\scriptsize\mbox{pot}} \label{use}
\\
&&= \frac{1}{2P^+}\lim_{\Delta \to 0}
  \Bigl\{ \frac{\partial}{i\partial \Delta^i} \langle P'S'| \bar{\psi}(0)
 \gamma^+
\int dy^- {\mathcal K}(y^- ) \, {\mathcal W}^-_{0y}\, gF^{+j}(y^-) {\mathcal W}^-_{y0} \psi(0)|PS\rangle
 -(i\leftrightarrow j) \Bigr\}\,. \nonumber
\eeq
 Parity and time--reversal ($PT$) symmetry tells that the following parametrization of the matrix element is possible
 \beq
\langle P'S'| \bar{\psi}(0)
 \gamma^+
\int dy^- {\mathcal K}(y^- ) \, {\mathcal W}^-_{0y}\, gF^{+i}(y^-) {\mathcal W}^-_{y0} \psi(0)|PS\rangle
=i\epsilon^{ij} \Delta_j  \bar{S}^+ h(\xi) + \cdots\,, \label{th}
\eeq
where $\bar{S}=(S+S')/2$ and $\xi \equiv -\Delta^+/2\bar{P}^+$ is the skewness parameter. [The dependence of $h(\xi)$ on the renormalization scale is implicit. We also suppress the dependence on $\Delta^2 \approx -\Delta_T^2$ since it is of higher order. Similar comments apply to other distributions defined below.]
This leads to
\beq
L_{\scriptsize\mbox{pot}}=  h(0)
\frac{S^+}{P^+}\,.
\eeq
 (\ref{th}) thus defines the potential angular momentum as the matrix element of a manifestly gauge invariant operator.\\

 The quark--gluon mixed operator that appears in the matrix element (\ref{th}) is familiar in the context of the twist--three mechanism of the single  spin asymmetry (SSA). Let us pursue this analogy and consider the following non-forward matrix element
\beq
T^{\mu\nu}(x_1,x_2,\xi) &=& \int \frac{dy^- dz^-}{(2\pi)^2} e^{\frac{i}{2}(x_1+x_2)\bar{P}^+z^- + i(x_2-x_1)\bar{P}^+y^-} \nonumber \\
 && \qquad \qquad \times \langle P'S'| \bar{\psi}(-z^-/2)
 \gamma^+
{\mathcal W}^-_{\frac{-z}{2}y}\, gF^{\mu\nu}(y^-) {\mathcal W}^-_{y\frac{z}{2}} \psi(z^-/2)|PS\rangle \nonumber \\
 &=& \frac{1}{\bar{P}^+}\epsilon^{\mu\nu\rho \sigma}\bar{S}_\rho \bar{P}_\sigma \Psi(x_1,x_2,\xi) + \frac{1}{\bar{P}^+} \epsilon^{\mu\nu\rho\sigma}\bar{S}_\rho \Delta_\sigma \Phi(x_1,x_2,\xi)+ \cdots\,. \label{nonf}
\eeq
  By symmetry considerations, it follows that $\Psi(x_1,x_2,\xi) = \Psi(x_2,x_1,-\xi)$ and $\Phi(x_1,x_2,\xi)=-\Phi(x_2,x_1,-\xi)$.   In the forward limit, and in the transversely polarized case $S^\mu = \delta^\mu_i S^i$, only the $\Psi$--term survives. The function $\Psi(x_1,x_2,0)$ plays the cental role in the so--called soft gluonic pole mechanism of the SSA \cite{Qiu:1991wg,Koike:2006qv}.
 In the longitudinally polarized case $\bar{S}^\mu\approx \delta^\mu_+ \bar{S}^+$, the $\Psi$--term vanishes for the relevant component $\mu\nu=+j$.
 By performing  Fourier transformations, we find
\beq
&& \langle P'S'| \bar{\psi}(0)
 \gamma^+
\int dy^- {\mathcal K}(y^- )\, {\mathcal W}_{0y} \, g F^{+j}(y^-) {\mathcal W}^-_{y0} \psi(0)|PS\rangle
\nonumber \\ &&\qquad = i\bar{P}^+\int dx_1 dx_2\, {\mathcal K}(x_1-x_2) T^{+j}(x_1,x_2) \nonumber \\
 && \qquad = i\epsilon^{jk}\bar{S}^+ \Delta_k \int dXdx\, {\mathcal K}(x) \Phi(X,x,\xi)\,,
\eeq
 where we switched to the notation $X=\frac{x_1+x_2}{2}$, $x=x_1-x_2$. The kernel is
\beq
{\mathcal K}(x) = \mbox{p.v.}\frac{1}{x} = \frac{1}{2}\left(\frac{1}{x+i\epsilon} + \frac{1}{x-i\epsilon} \right)\,,
\eeq
 in the case ${\mathcal K}(y^-)=\frac{1}{2}\epsilon(y^-)$ and
 \beq
 {\mathcal K}(x) = \frac{1}{x \pm i\epsilon}\,,
 \eeq
 in the cases ${\mathcal K}(y^-) = \pm \theta(\pm y^-)$.
Comparing with (\ref{th}), we obtain an alternative expression for the potential angular momentum
\beq
L_{\scriptsize\mbox{pot}}=\int dX dx\, {\mathcal K}(x) \Phi(X,x,0)\,.
\eeq
Note that, since $\Phi(X,0,0)=0$, different choices for ${\mathcal K}$ lead to the same result, as they should.

\section{Canonical angular momentum}

Next we exploit the relation between the twist--three approach to the SSA and the approach based on the transverse momentum dependent distribution (TMD).\footnote{The following discussion is similar to the works of Ref.~\cite{Hagler:2003jw,Lorce:2011kd}. We improve upon these works by fully taking into account the gauge field and the issue of gauge invariance.}
 In the longitudinally polarized and non-forward case, we define
\beq
&&f(x,q_T,\Delta) \equiv \int \frac{dz^- d^2z_T}{(2\pi)^3} e^{ix\bar{P}^+z^- -iq_T \cdot z_T}
  \label{bou} \\
&& \qquad \times \langle P'S'|\bar{\psi}(-z^-/2,-z_T/2)\gamma^+ {\mathcal W}^-_{\frac{-z}{2},\pm \infty}{\mathcal W}^T_{\frac{-z_T}{2}, \frac{z_T}{2}}{\mathcal W}^-_{\pm \infty, \frac{z}{2}}\psi(z^-/2,z_T/2)|PS\rangle\,, \nonumber
\eeq
 where ${\mathcal W}^T$ is the Wilson line in the transverse direction at $x^-=\pm \infty$.
 In the forward case $\Delta=0$, the matrix element (\ref{bou}) reduces to the usual TMD.  As is well--known in that context, there is freedom in choosing the path connecting the points $(-z^-/2,-z_T/2) \to (z^-/2,z_T/2)$. Using the Wilson line that goes to future infinity and then comes back $-z^-/2\to +\infty \to z^-/2$, one takes care of the final state interaction. The TMD in this case is relevant to the semi-inclusive DIS (SIDIS). The other case $-z^-/2 \to -\infty \to z^-/2$ includes the initial state interaction relevant to the Drell--Yan process. For the present purpose, one may as well take the average of the two cases.

The relation between (\ref{bou}) and (\ref{th}) is revealed by taking the second moment of $f$  in $q_T$ \cite{Boer:2003cm}\footnote{Cf. Eq.~(39) of \cite{Boer:2003cm}. Via partial integration, $q_T^i$ is replaced by the spatial derivative $\partial^i_T$. When acting on the Wilson line, it brings down the factor $\partial^i_T A^+ =D^+A^i+F^{i+}$ which reduces to the two terms in (\ref{vani}).}
\beq
&& F^i(x,\Delta) \equiv  \int d^2q_T\, q_T^i f(x,q_T,\Delta) \nonumber  \\
&& \qquad
 =\frac{1}{2}\int \frac{dz^-}{2\pi}e^{ix\bar{P}^+z^-} \Biggl\{ \langle P'S'|\bar{\psi}(-z^-/2)\gamma^+ \left({\mathcal W}^-_{\frac{-z}{2},\frac{z}{2}} i\overrightarrow{D}^i - i\overleftarrow{D}^i{\mathcal W}^-_{\frac{-z}{2},\frac{z}{2}}\right)\psi(z^-/2)|PS\rangle
\nonumber \\
 && \qquad \qquad \qquad  \qquad - \langle P'S'| \bar{\psi}(-z^-/2)\gamma^+ \int dy^-\left({\mathcal K}(y^- -z^-/2)+{\mathcal K}(y^- + z^-/2)\right) \nonumber \\
 && \qquad \qquad \qquad \qquad \qquad \qquad \qquad \times {\mathcal W}^-_{\frac{-z}{2},y}\, gF^{+i}(y^-){\mathcal W}^-_{y,\frac{z}{2}}\psi(z^-/2)|PS\rangle \Biggr\}\,,  \label{vani}
 \eeq
 where the kernel ${\mathcal K}$ is in one--to--one correspondence with the choice of the Wilson line path in (\ref{bou}). In the forward case, (\ref{vani}) vanishes for the longitudinal polarization by rotational symmetry in the transverse plane. In the non-forward case, however, the following structure
 \beq
 f(x,q_T,\Delta)\sim \frac{i}{\bar{P}^+}\epsilon^{+-ij}\bar{S}^+q_{Ti} \Delta_j \tilde{f}(x,q_T^2,\xi,\Delta_T\cdot q_T)\,,
 \eeq
  is allowed, so that (\ref{vani}) is not necessarily zero.  Note that, because of an extra minus sign from $\Delta_j$ under the $PT$ transformation, the function $\tilde{f}$ does {\it not} change signs when changing the directions of the Wilson line,\footnote{More precisely, $\tilde{f}(x,q_T^2,\xi,\Delta_T\cdot q_T) \to +\tilde{f}(x,q_T^2,-\xi,-\Delta_T\cdot q_T)$ after changing the directions. But the difference is immaterial  in the limit $\Delta \to 0$.}  in contrast to the known sign flip of the spin--dependent TMDs in the SIDIS and Drell--Yan reactions \cite{Collins:2002kn}.

 We then take the first  moment in $x$
\beq
&& \int dx F^i(x,\Delta) =
\frac{1}{\bar{P}^+} \Bigl\{ \frac{1}{2}\langle P'S'|\bar{\psi}(0)\gamma^+ (i\overrightarrow{D}^i -i\overleftarrow{D}^i)\psi(0)|PS\rangle
\nonumber \\
 && \qquad \qquad \qquad \qquad - \langle P'S'| \bar{\psi}(0)\gamma^+ \int dy^-{\mathcal K}(y^- ){\mathcal W}^-_{0y}\, gF^{+i}(y^-){\mathcal W}^-_{y0}\psi(0)|PS\rangle \Bigr\} \nonumber \\
 && \qquad \qquad = \frac{1}{\bar{P}^+} \Bigl\{ \frac{1}{2}\langle P'S'|\bar{\psi}(0)\gamma^+ (i\overrightarrow{D}^i -i\overleftarrow{D}^i )\psi(0)|PS\rangle
+ \langle P'S'| \bar{\psi}(0)\gamma^+A^i_{\scriptsize\mbox{phys}} \psi(0)|PS\rangle \Bigr\}
 \nonumber \\ && \qquad \qquad = \frac{1}{2\bar{P}^+}  \langle P'S'|\bar{\psi}(0)\gamma^+ (i\overrightarrow{D}^i_{\scriptsize\mbox{pure}} -i\overleftarrow{D}^i_{\scriptsize\mbox{pure}} )\psi(0)|PS\rangle\,.
 \eeq
The matrix element (\ref{th}) indeed appears in the second term of the first equality, but somewhat remarkably, it is absorbed by the covariant derivative in the first term. The final expression features precisely the `pure gauge' part of the covariant derivative $D_{\scriptsize\mbox{pure}}$.
  Differentiating with respect to $\Delta$, we arrive at
 \beq
 \epsilon^{ij}L_{\scriptsize\mbox{can}} &=& \frac{1}{2P^+}\lim_{\Delta \to 0} \frac{\partial}{i\partial \Delta^i }\langle P'S'|\bar{\psi}(0)\gamma^+ (i\overrightarrow{D}^j_{\scriptsize\mbox{pure}} -i\overleftarrow{D}^j_{\scriptsize\mbox{pure}})\psi(0)|PS\rangle \nonumber \\
 &=&  \lim_{\Delta \to 0} \frac{\partial}{i\partial   \Delta^i } \int dx F^j(x,\Delta)
 \nonumber \\
 &=&  \lim_{\Delta \to 0} \frac{\partial}{i\partial \Delta^i } \int dx  d^2q_T\,q_T^j f(x,q_T,\Delta)  \nonumber \\
 &=& \epsilon^{ij}\frac{S^+}{P^+}   \frac{1}{2}\int dx d^2q_T\, q_T^2 \tilde{f}(x,q_T^2)\,.
  \label{canoni}
  \eeq
  This is a formula relating the canonical OAM of quarks to the matrix element of a well--defined, manifestly gauge invariant operator.\footnote{We note that a possible UV regularization of the $q_T$--integral (\ref{canoni}) and its evolution require an entirely separate analysis.} The final expression in (\ref{canoni}) agrees with
  the OAM constructed by Lorc\'e and Pasquini \cite{Lorce:2011kd} from the Wigner distribution (neglecting gauge invariance). We have thus established a gauge invariant link between the Wigner distribution approach and the spin decomposition framework of Chen {\it et al.}

 Similarly, for the gluon orbital angular momentum one can define
\beq
&&g(x,q_T,\Delta) \equiv -i \int \frac{dz^- d^2z_T}{(2\pi)^3} e^{ix\bar{P}^+z^- -iq_T \cdot z_T}
 \label{oamg} \\
&& \qquad \quad \times \langle P'S'|F^{+\alpha}(-z^-/2,-z_T/2) {\mathcal W}^-_{\frac{-z}{2},\pm \infty}{\mathcal W}^T_{\frac{-z_T}{2}, \frac{z_T}{2}}{\mathcal W}^-_{\pm \infty, \frac{z}{2}}A_{\alpha}^{\scriptsize\mbox{phys}}(z^-/2,z_T/2)|PS\rangle\,, \nonumber
\eeq
 where $A_{\scriptsize\mbox{phys}}$ is as in (\ref{hatta}), and  now the Wilson lines are in the adjoint representation. Under the $PT$ transformation, one gets the same operator back (up to the direction of the Wilson line) provided the skewness parameter $\xi$ vanishes, which we assume here.   In deeply virtual Compton scattering (DVCS), $\xi=0$ corresponds to the elastic scattering of the photon.

 Proceeding as before, one finds the double moment in $x$ and $q_T$
 \beq
\int dx  d^2q_T\, q_T^i g(x,q_T,\Delta) =  \frac{1}{2\bar{P}^+}  \langle P'S'|F^{+\alpha} (\overrightarrow{D}^i_{\scriptsize\mbox{pure}} -\overleftarrow{D}^i_{\scriptsize\mbox{pure}} )A_{\alpha}^{\scriptsize\mbox{phys}}|PS\rangle\,.
 \eeq
The matrix element of  (\ref{26}) thus becomes
\beq
\frac{1}{2P^+} \frac{\langle PS| \int dx^- d^2x_T\, M_{\scriptsize\mbox{gluon-orbit}}^{+ij}|PS\rangle}{(2\pi)^3\delta^3(0)}&=&  \lim_{\Delta \to 0} \frac{\partial}{i\partial \Delta^i } \int dx  d^2q_T\,q_T^j g(x,q_T,\Delta) \nonumber \\
&=& \epsilon^{ij} \frac{S^+}{P^+}\frac{1}{2}\int dx d^2q_T \, q_T^2 \tilde{g}(x,q_T^2)  \,, \label{glu}
\eeq
 where we parameterized, as $\Delta \to 0$,
 \beq
  g(x,q_T,\Delta)= \frac{i}{\bar{P}^+}\epsilon^{+-ij}\bar{S}^+q_{Ti} \Delta_j \tilde{g}(x,q_T^2) + \cdots\,.
\eeq

\section{Conclusions}
 We have shown that the physical part of the gauge field $A_{\scriptsize\mbox{phys}}$ proposed in \cite{Hatta:2011zs}  leads to well--defined expressions for the canonical and potential angular momenta  in terms of the matrix element of certain gauge invariant operators. If one defines $\Delta G$ as the gluon helicity, consistency requires that (\ref{canoni}) is the  corresponding canonical angular momentum.  Now that we can, at least in principle, measure $L_{\scriptsize\mbox{Ji}}$, $L_{\scriptsize\mbox{can}}$ and the difference $L_{\scriptsize\mbox{pot}}=L_{\scriptsize\mbox{Ji}}-L_{\scriptsize\mbox{can}}$  separately,
 it seems more legitimate to call $L_{\scriptsize\mbox{can}}$, or equivalently the OAM from the Wigner distribution \cite{Lorce:2011kd,Lorce':2011ni}, the quarks' genuine orbital angular momentum since it satisfies the fundamental commutation relation.

Regarding measurability, it should be possible to compute (\ref{canoni}) in lattice QCD simulations as in the case of the ordinary (forward) TMD (see, e.g., \cite{Musch:2010ka}).
 Since there is no sign flip in (\ref{bou}) when changing the directions of the Wilson lines, the matrix element may not be very sensitive to the choice of the path. If so, and if one is not interested in the $x$--dependence, one may first integrate over $x$ in (\ref{bou}) and connect the quark operators (at the same value of $z^-=0$) by a purely spatial Wilson line in the transverse plane. This avoids the introduction of lightlike Wilson lines on a Euclidean lattice which appears to be a vague issue. For the gluon orbital angular momentum (\ref{glu}) with $g$ defined in (\ref{oamg}), one has to deal with lightlike Wilson lines even after the $x$--integration.

Finally, from the experimental point of view, the matrix elements such as (\ref{nonf}) are related to the twist--three GPDs \cite{Anikin:2000em,Belitsky:2000vx} which are hard to extract. It would be interesting to see if there are  processes in which these functions contribute  to the cross section at leading order as in the single spin asymmetry.\\

 {\it Acknowledgments}---I am indebted to Kazuhiro Tanaka for many discussions, and for reading the manuscript and bringing \cite{Hagler:2003jw} to my attention. I also thank Terry Goldman, Yuji Koike and Jian-Wei Qiu for helpful conversations and correspondence.
 This work is supported by Special Coordination Funds for Promoting Science and Technology
 of the Ministry of Education, the Japanese Government.

\end{document}